\pdfoutput=1
\documentclass[11pt,twoside,a4paper,cmspaper,final,collab]{cms-tdr}
\def\svnVersion{af44e50}\def\svnDate{2024/01/03}\def\cmsCernNoTag{CERN-EP-2023-165}\def\cmsCernDate{\today}\def\cmsMessage{Published in the Journal of High Energy Physics as \href{http://dx.doi.org/10.1007/JHEP12(2023)070}{\doi{10.1007/JHEP12(2023)070}.}}
\begin{document}\cmsNoteHeader{EXO-21-005}

\newcommand{\epsSq}{\ensuremath{\varepsilon^2}\xspace}
\newcommand{\thetaH}{\ensuremath{\theta_{\PH}}\xspace}
\newcommand{\Lxy}{\ensuremath{L_{\mathrm{xy}}}\xspace}
\newcommand{\sigLxy}{\ensuremath{{\sigma}_{L_{\mathrm{xy}}}}\xspace}
\newcommand{\Mmumu}{\ensuremath{m_{\PGm\PGm}}\xspace}
\newcommand{\DYTurbo}{\textsc{DYTurbo}\xspace}
\newcommand{\Zdark}{\ensuremath{\PZ_{\mathrm{D}}}\xspace}

\cmsNoteHeader{EXO-21-005}

\title{Search for direct production of GeV-scale resonances decaying to a pair of muons in proton-proton collisions at \texorpdfstring{$\sqrt{s} = 13\TeV$}{sqrt(s) = 13 TeV}}

\date{\today}

\abstract{
A search for direct production of low-mass dimuon resonances is performed using $\sqrt{s}=13\TeV$ proton-proton collision data collected by the CMS experiment during the 2017--2018 operation of the CERN LHC with an integrated luminosity of $96.6\fbinv$. The search exploits a dedicated high-rate trigger stream that records events with two muons with transverse momenta as low as 3\GeV but does not include the full event information. The search is performed by looking for narrow peaks in the dimuon mass spectrum in the ranges of 1.1--2.6\GeV and 4.2--7.9\GeV. No significant excess of events above the expectation from the standard model background is observed.  Model-independent limits on production rates of dimuon resonances within the experimental fiducial acceptance are set. Competitive or world's best limits are set at 90\% confidence level for a minimal dark photon model and for a scenario with two Higgs doublets and an extra complex scalar singlet (2HDM+S). Values of the squared kinetic mixing coefficient \epsSq in the dark photon model above $10^{-6}$ are excluded over most of the mass range of the search. In the 2HDM+S, values of the mixing angle $\sin(\thetaH)$ above 0.08 are excluded over most of the mass range of the search with a fixed ratio of the Higgs doublets vacuum expectation $\tan\beta=0.5$. 
}

\hypersetup{
pdfauthor={CMS Collaboration},
pdftitle={Search for prompt production of GeV-scale resonances decaying to a pair of muons in proton-proton collisions at sqrt(s) = 13 GeV},
pdfsubject={CMS},
pdfkeywords={CMS, hadron-hadron scattering, beyond the standard model}}
\maketitle

\section{Introduction\label{sec:introduction}}

Searches for resonant particle pair production in accelerator experiments have played a significant role in the development of the standard model (SM). The observation of resonant production of electron-positron pairs and muon-antimuon pairs signaled the discovery of the charm~\cite{PhysRevLett.33.1404,PhysRevLett.33.1406} and bottom~\cite{PhysRevLett.39.252} quarks, respectively. The \PZ boson was discovered through the observation of resonant production of both electron-positron and muon-antimuon pairs~\cite{1983398}. More recently, findings of resonant production of photon pairs and \PZ boson pairs were central to the discovery of the Higgs boson at the CERN LHC by the ATLAS~\cite{Aad:2012tfa} and CMS~\cite{Chatrchyan:2012ufa,Chatrchyan:2013} Collaborations. Searches for new particles at the \TeVns scale are an important component of the LHC physics program. Resonant searches in dilepton channels are a large part of this effort, however, they have so far yielded results consistent with SM predictions~\cite{Aad:2019fac,Sirunyan:2021khd}. Interest has therefore grown in extending resonance searches to explore the phase space of lower masses and very small effective couplings to SM particles, which is becoming possible thanks to the large amount of data delivered by the LHC. The \GeVns scale is particularly interesting because a new vector boson in that mass range could provide an explanation for several astrophysical observations~\cite{ArkaniHamed:2008qn}.

This paper describes a search for production of a narrow resonance that decays promptly to a pair of oppositely charged muons using proton-proton ($\Pp\Pp$) collision data recorded by the CMS experiment. The explored mass ranges are 1.1--2.6\GeV and 4.2--7.9\GeV. These ranges exclude the regions around the $\PJGy$, $\Pgy$, and $\PGUP{1S}$ resonances. Previous limits on dimuon resonance production in these mass ranges have been set by the BaBar~\cite{Lees:2014xha} and LHCb~\cite{Aaij:2017rft,Aaij:2019bvg,Aaij:2020ikh} Collaborations. The latter also probes the mass range below 1\GeV and searches for long-lived particles, which are not covered in this paper. This search is an extension of a previous CMS analysis for higher masses~\cite{Sirunyan:2019wqq}, with trigger selection and muon identification optimized for the very low-mass region. It assumes that the natural width of the resonance is much smaller than the detector dimuon mass resolution ($\approx$1.3\%), and that its decay is consistent with originating from the primary $\Pp\Pp$ collision vertex within the instrumental resolution. The sensitivity of this analysis is enhanced by the use of high-rate (``scouting'') triggers~\cite{CMS:scouting} that select a larger fraction of signal-like events as compared with the standard CMS triggers, by recording only a small part of the event information, consisting of trigger objects as opposed to the full detector readout. The previous CMS analysis at higher masses utilized both the standard CMS triggers and the scouting triggers, while this search uses the scouting triggers exclusively.

\begin{figure}[!htb]
{\centering
\includegraphics[width=0.40\linewidth]{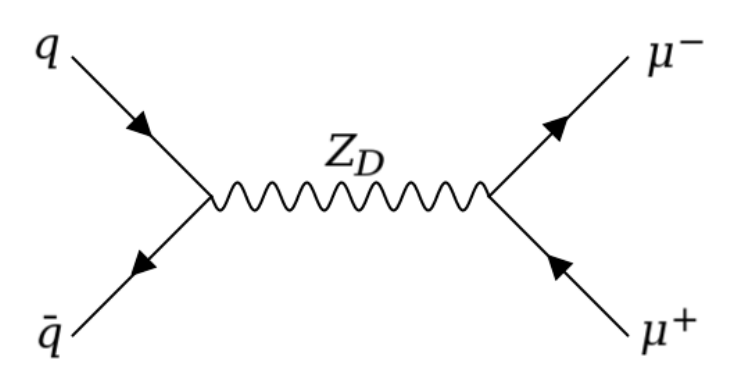}
\includegraphics[width=0.40\linewidth]{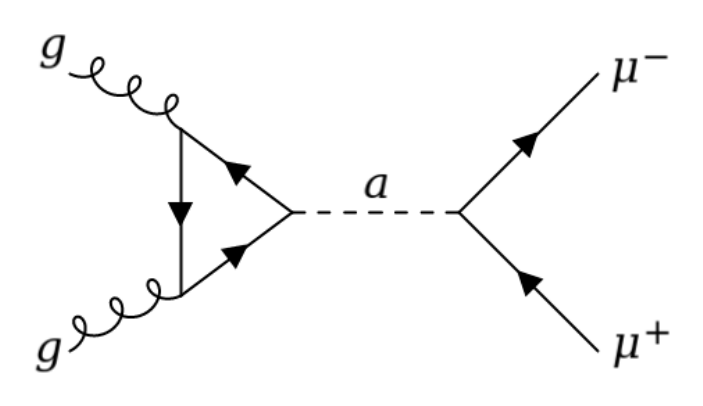}
\caption{Expected dominant production modes for beyond-SM particles: a dark photon $\Zdark$ (left) and light pseudoscalar boson \Pa (right).}
\label{fig:diagrams}
\par}
\end{figure}

We also provide an interpretation of the result in the context of two specific beyond the SM scenarios. The first scenario involves the introduction of a new gauge field $U(1)_\text{D}$ that undergoes kinetic mixing with the SM hypercharge gauge field $U(1)_Y$~\cite{GALISON1984279,HOLDOM1986196}. 
Such mixing, controlled by the parameter $\varepsilon$, results in a dark photon ($\Zdark$), a possibly massive, neutral vector boson. In the mass ranges probed by this search, it is coupled dominantly to the electric charges of SM particles with a coupling strength suppressed by a factor $\varepsilon$.
It is assumed to be produced via the Drell--Yan (DY) mechanism, which is the dominant production mechanism in the mass range of interest~\cite{hahm1}. The second scenario is a two Higgs doublet model with an extra complex scalar singlet (2HDM+S)~\cite{Haisch:2018kqx}. 
For certain model parameters~\cite{Argyropoulos:2022ezr}, this scenario can accommodate a light pseudoscalar boson (\Pa) that can decay into a dimuon pair.
In this model, the couplings of the light pseudoscalar \Pa to the SM particles are determined by its mixing with the Higgs doublets and parameterized by the mixing angle ($\thetaH$) and the ratio of the Higgs doublet vacuum expectation values (tan$\beta$). It is assumed that the light pseudoscalar \Pa is predominantly produced via gluon-gluon fusion (ggF). The interpretation is provided in the specific 2HDM+S scenario of type IV with tan$\beta=0.5$ given the enhanced coupling to the up-type quarks, predicted in this model. The same model is targeted by the LHCb experiment~\cite{Aaij:2020ikh} because it represents the most sensitive one in experiments at pp colliders and allows meaningful bounds to be set on the sine of the mixing angle $\thetaH$.  Example Feynman diagrams for the production of both particles are presented in Fig.~\ref{fig:diagrams}. 

This paper is organized as follows.  We begin with the CMS detector description in Section~\ref{sec:cms} and elaborate on the relevant triggers (Section~\ref{sec:trigger}) and the simulated data samples used in the analysis (Section~\ref{sec:simulation}). Sections~\ref{sec:selection} and~\ref{sec:modelfit} detail the event selection and the signal and background modeling, respectively. Section~\ref{sec:uncertainty} explains the systematic uncertainties and the results are presented in Section~\ref{sec:limit}, followed by a summary in Section~\ref{sec:summary}. Tabulated results are provided in the HEPData record for this analysis~\cite{hepdata}.
\section{The CMS detector\label{sec:cms}}

The central feature of the CMS apparatus is a superconducting solenoid of 6\unit{m} internal diameter, providing a magnetic field of 3.8\unit{T}. Within the solenoid volume are a silicon pixel and strip tracker, and both a lead tungstate crystal electromagnetic and a brass/scintillator calorimeter, which consist of a barrel and two endcap sections. Forward calorimeters extend the pseudorapidity $\eta$ coverage provided by the barrel and endcap detectors. Muons are detected in gas-ionization chambers embedded in the steel flux-return yoke outside the solenoid.  A more detailed description of the CMS detector, together with a definition of the coordinate system used and the relevant kinematic variables, is given in Ref.~\cite{Chatrchyan:2008zzk}.

Muons are measured in the pseudorapidity range $\abs{\eta} < 2.4$, with detection planes made using three technologies: drift tubes, cathode strip chambers, and resistive plate chambers. Matching muons to tracks measured in the silicon tracker results in a relative transverse momentum (\pt) resolution of 1\% in the barrel and 3\% in the endcaps, for muons with \pt up to 100\GeV.
\section{Triggers and data sets\label{sec:trigger}}

Events of interest are selected using a two-tiered trigger system~\cite{CMS_trigger}, the level-1 (L1) and high-level (HLT) triggers.  The L1, composed of custom hardware processors, uses information from the calorimeters and muon detectors to select events at a rate of around 100\unit{kHz}.  The HLT consists of a farm of CPU processors running a version of the CMS full event reconstruction software optimized for fast event data processing.  Events passing the full set of HLT selection criteria are saved for permanent storage at an overall rate of about 1\unit{kHz}.

A set of four different L1 trigger algorithms are utilized in this analysis, defined by requiring the presence of at least two muons with various additional requirements. The simplest of these (1) only requires two muons with $\pt>15\GeV$ and $>7\GeV$, respectively. The remaining three triggers all share the requirement that the two muons be oppositely charged, together with the additional conditions: (2) each muon has $\abs{\eta}<2.0$ and $\pt>4.5\GeV$, and that the dimuon mass ($\Mmumu$) is between 7--18\GeV; or (3) each muon has $\pt>4.0 \ (4.5)\GeV$ in 2017 (2018) and the two muons must have an angular separation $\DR =\sqrt{\smash[b]{(\Delta\eta)^2+(\Delta\phi)^2}}<1.2$ and $\phi$ is the azimuthal angle; or (4) both muons have $\abs{\eta}<1.5$ and are separated by $\Delta R<1.4$, with no restriction on \pt.  These criteria are summarized in Table~\ref{tab:L1}.

\begin{table}[htb]
\topcaption{The set of dimuon L1 requirements applied in the high-rate triggers. }
\def\arraystretch{1.3}
\label{tab:L1}
\centering\begin{tabular}{ccccccc}
\hline
L1 path & \pt [\GeVns{}] & $\abs{\eta}$ & $\Delta R$ & $\Mmumu$ [\GeVns{}] & Charge \\
\hline
(1)     & $>$15/7  & \NA           & \NA         & \NA                & \NA \\
(2)     & $>$4.5   & $<$2.0      & \NA         & 7--18             & Opposite sign \\
(3)     & $>$4.0 (2017)/4.5 (2018) & \NA           & $<$1.2    & \NA                & Opposite sign \\
(4)     & \NA        & $<$1.5      & $<$1.4    & \NA                & Opposite sign \\
\hline
\end{tabular}
\end{table}

\begin{figure}[!htb]
{\centering
\includegraphics[width=0.49\linewidth]{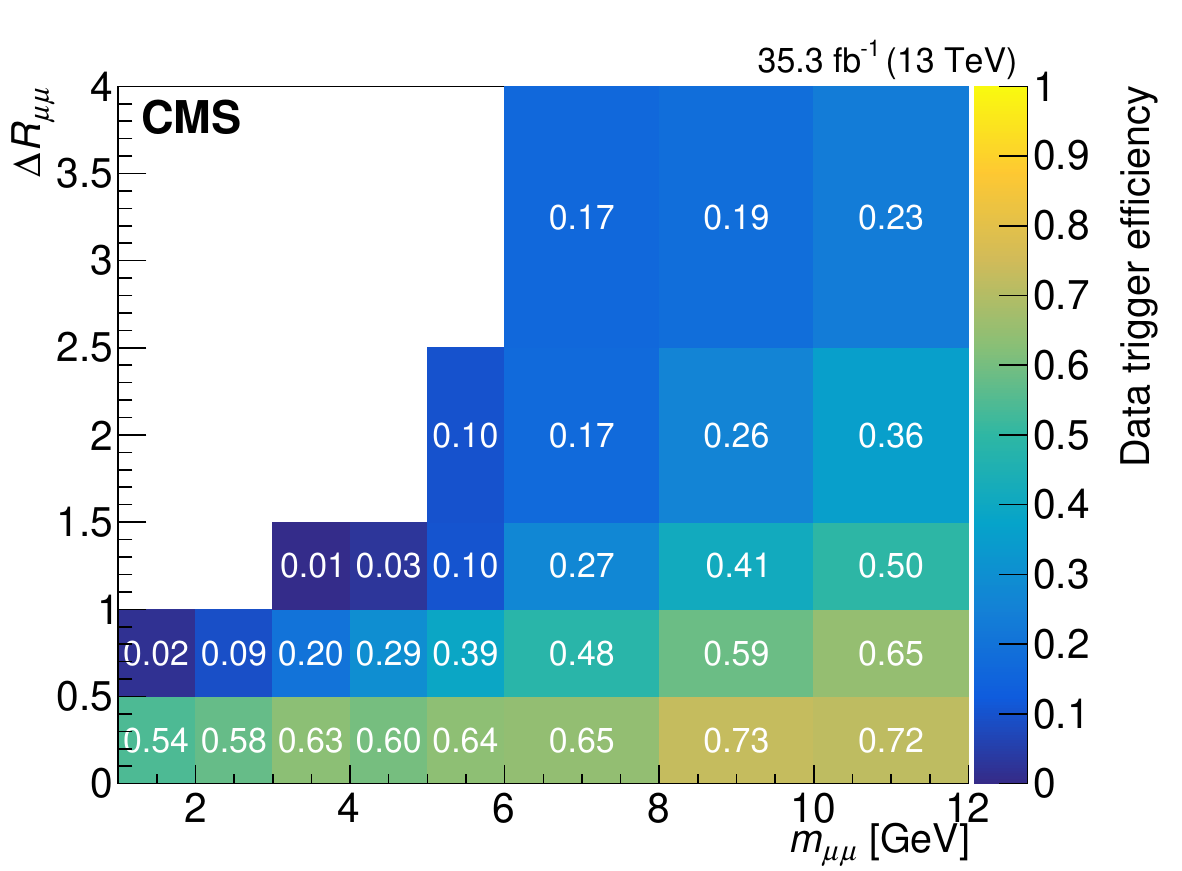}
\includegraphics[width=0.49\linewidth]{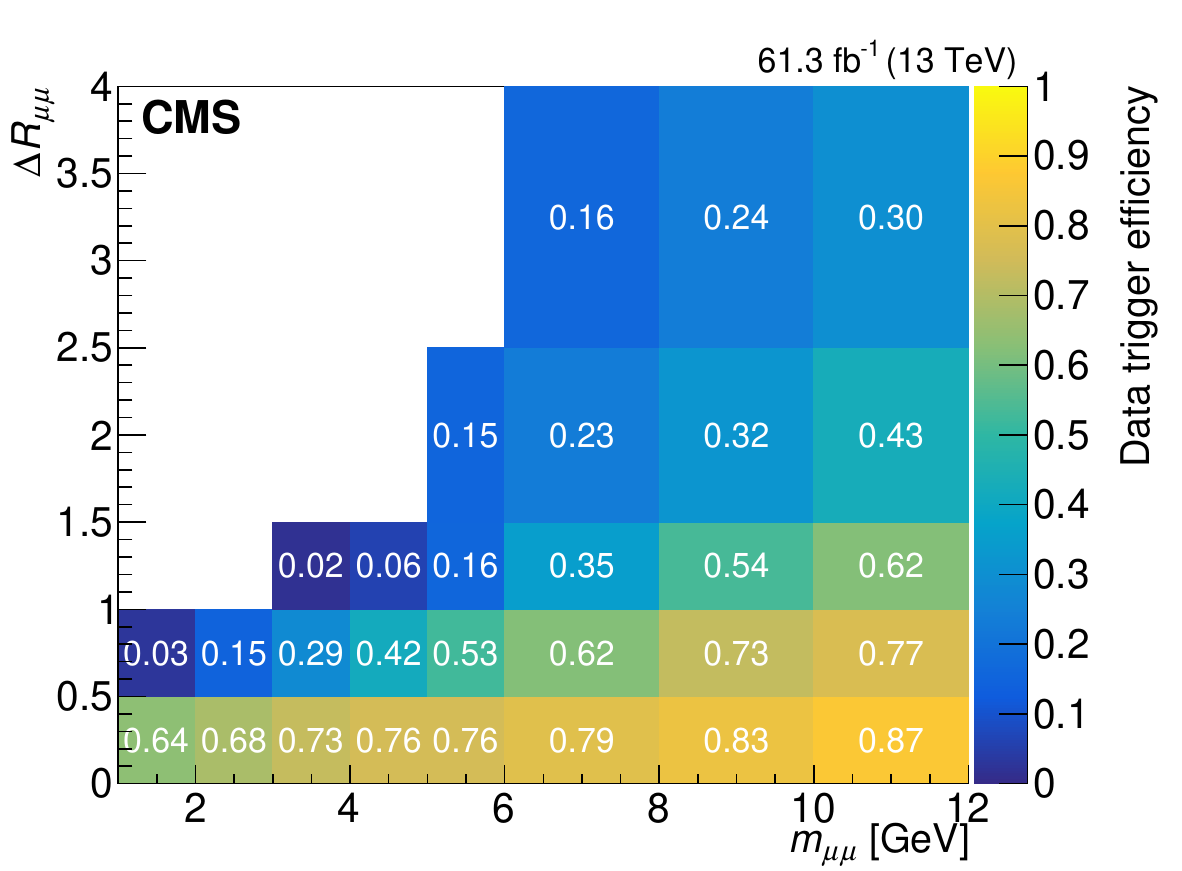}
\caption{The measured 2017 (left) and 2018 (right) efficiencies of the dimuon scouting trigger and logical OR of all L1 triggers using 2017--2018 data.  Each cell value represents the probability that a pair of muons meeting the trigger requirements will activate the dimuon scouting trigger. The $x$-axis shows the dimuon mass and includes the entire relevant range for this analysis.  The $y$-axis shows the angular separation, $\Delta R$, between the two muons.  Statistical uncertainty on the value of each cell is less than 5\%.}
\label{fig:trigEffs}
\par}
\end{figure}

At the HLT, our analysis makes use of the CMS high-rate data scouting trigger system~\cite{CMS:scouting} in which the information associated with the selected events is much reduced, permitting the recording of larger data samples. This technique has been used in previous resonant searches performed by the CMS experiment including dimuon~\cite{Sirunyan:2019wqq} and hadronic~\cite{ref:hadronicScouting} final states.
Events are directly reconstructed in the HLT and are required to have at least two muons with $\pt>3\GeV$. The dimuon events selected by the data scouting trigger are written to disk with an event size of about 4~(8)\unit{kB} for data taken in 2017~(2018).  The stored information includes tracking hits, calorimeter clusters, muon hits, and trigger primitives, along with their reconstructed quantities.  By contrast the size of a standard raw event is closer to 1\unit{MB}. These dimuon scouting triggers were fully commissioned during 2017~\cite{Mukherjee:2017wcl} and recorded events at a rate of approximately 2\unit{kHz} at the peak instantaneous luminosity of $2\times 10^{34} \ \mathrm{cm}^{-2}\mathrm{s}^{-1}$, whereas the standard dimuon triggers recorded approximately 0.45\unit{kHz} at the same instantaneous luminosity. Data corresponding to an integrated luminosity of 96.6\fbinv were collected with these triggers in 2017--2018. Figure~\ref{fig:trigEffs} shows the efficiency of the dimuon scouting triggers in conjunction with the four L1 triggers for both data periods.  
This is measured using an independent CMS data set collected with a trigger requiring large energy deposits in the electromagnetic calorimeter. This data set is agnostic to the presence or absence of muons in the event and therefore is not correlated with the activation of the dimuon scouting trigger. By sampling this data set for events that contain pairs of muons, we assemble an unbiased set of dimuon events to gauge the performance of the dimuon scouting trigger. For this sample of unbiased muon pairs, we count the number that also triggered the dimuon scouting path and at least one of the four L1 triggers, and compare the result to the total number of reconstructed dimuon events in that data set that pass our fiducial selections. The ratio of these two counts is the efficiency of our trigger.
The trigger efficiency in 2018 is higher than that of 2017, due to updates in the HLT muon reconstruction algorithm for the 2018 data and efficiency loss in the later part of 2017 caused by a powering problem affecting part of the pixel detector~\cite{Phase1Pixel}.

\section{Event simulation\label{sec:simulation}}

The analysis employs an estimation of the SM background based on recorded data.  However, simulations of certain SM processes are used for the validation of selection criteria, corrections to data driven efficiency measurements, as well as the corresponding systematic uncertainties in the signal predictions.  These processes include DY production of dimuon pairs in the low $\Mmumu$ region ($1<\Mmumu<10\GeV$), as well as resonant dimuon production via $\PJGy$ and $\PGUP{1S}$ mesons.  These processes are simulated at leading order (LO) in perturbative quantum chromodynamics with $\PYTHIA$ 8.230~\cite{pythia}.  In the 2HDM+S scenario, the production of a BSM pseudoscalar Higgs via ggF, with modified couplings to the up-type and down-type quarks, is simulated at LO with $\PYTHIA$ 8.230 and is used as a proxy for the production of a light pseudoscalar of the type IV model. The $\Zdark$ signal is simulated at next-to-LO in perturbative quantum chromodynamics using the hidden Abelian Higgs model~\cite{hahm1,hahm2,hahm3} implemented in the $\MGvATNLO$ v3.4.1 generator~\cite{madgraph}, with up to two additional partons in the matrix element calculations. The parton distribution function (PDF) set is NNPDF3.1 in the four-flavor scheme~\cite{nnpdf}. 

For all simulated samples, the parton showering and hadronization processes are performed with $\PYTHIA$ 8.230 using the CP5 underlying event tune~\cite{cp5bib}.  Events are passed through a full detector simulation performed with $\GEANTfour$~\cite{geant4}.  
The simulated samples include additional pp interactions per bunch crossing, referred to as ``pileup''. The effect of pileup is taken into account by generating minimum-bias collision events with $\PYTHIA$. The simulated events are weighted such that the distribution of the number of pileup interactions matches that in data.

\section{Event selection\label{sec:selection}}

A complete event reconstruction is performed by the HLT.  The muons in the scouting data stream are reconstructed by combining the information from the silicon tracker and muon systems, and their momentum is determined from the curvature of the muon track.  
The muon momentum obtained with the HLT event reconstruction agrees with that obtained with the full CMS offline event reconstruction within 1--2\%.

The event candidate is required to have at least one primary vertex (PV) reconstructed by the HLT system and to contain a pair of oppositely charged muons originating from this vertex.  The absolute transverse distance of the PV to the beam axis, referred to as the vertex transverse displacement $\Lxy$, is required to be less than 0.2\unit{cm}. Both muons are required to have $\pt>4\GeV$ and $\abs{\eta}<1.9$.  The restriction on $\eta$ is imposed to ensure optimal dimuon mass resolution without incurring a significant loss in acceptance.  The muons are required to satisfy a muon identification based on multivariate analysis (MVA) techniques.  Two MVA discriminants are used depending on the reconstructed $\Mmumu$, optimized for the signal kinematic properties in each mass range.  The MVA identification utilizes information on the quality of the muon tracks, the relative isolation of the muon, and the vertex that the muons are associated with.  The muon track quality parameters include the number of hits in the pixel detector, number of contributing layers in the strip tracker, and $\chi^2$ of the muon track fit.  The relative isolation of the muon is computed as the scalar \pt sum of reconstructed tracks in a cone of $\DR = 0.3$ around the muon candidate, divided by the muon $\pt$.  The dimuon vertex information includes the $\chi^2$ of the common vertex fit and $\Lxy$. These parameters are combined into a single discriminator using a boosted decision tree (BDT)~\cite{bdt}.  The choice of input parameters to the BDT and the requirement on the discriminator are optimized to achieve the highest expected sensitivity to signal-like events.

The mass range 2.6--4.2\GeV is excluded because of the presence of the $\PJGy$ and $\Pgy$ resonances.  The MVA identification for $\Mmumu<2.6\GeV$ is trained using a partial $\PJGy$ sample, whereas the identification for $\Mmumu>4.2\GeV$ is trained using an $\PGUP{1S}$ sample.  The efficiency of the muon MVA identification is determined from $\PJGy$ and $\PGU$ events using the ``tag-and-probe technique''~\cite{ref:TnP}.  Background samples enriched with nonprompt and misidentified muons for the BDT training are extracted from the data events with only one pair of same-sign muons and a mass of $3<\Mmumu<10\GeV$.

\begin{figure}[!htb]
{\centering
\includegraphics[width=0.8\linewidth]{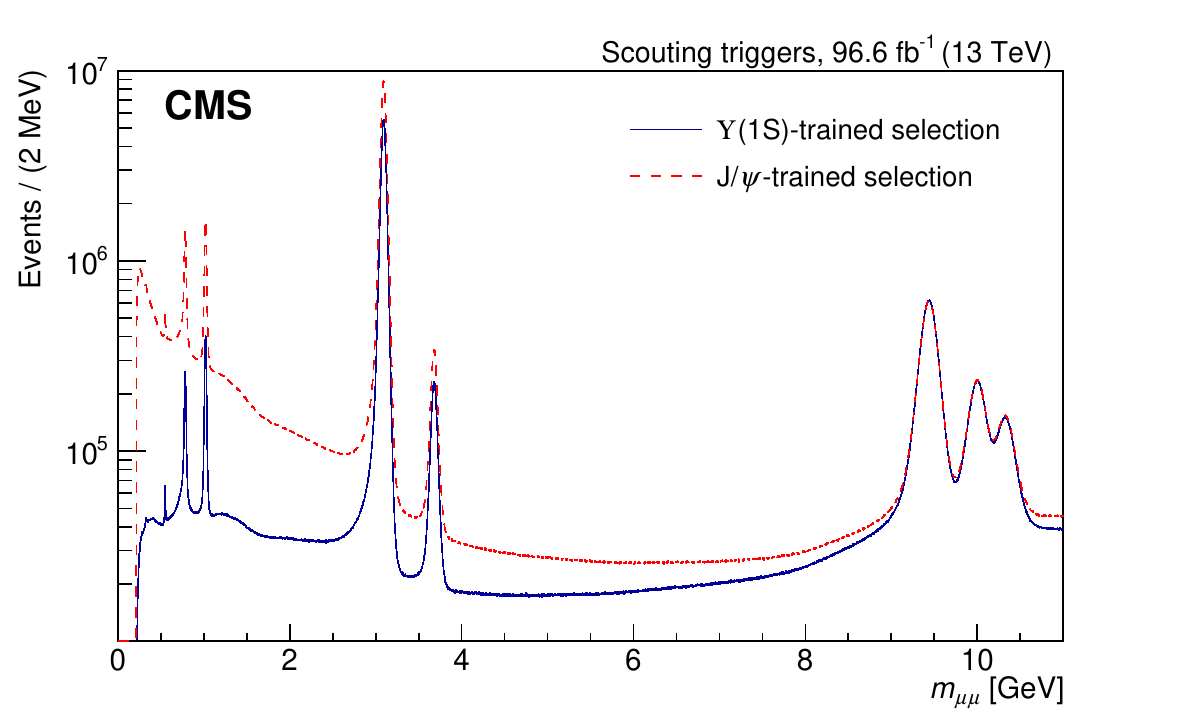}
\caption{The $\Mmumu$ distribution obtained with the muon scouting data collected during 2017--2018 with two sets of selections: the $\PGUP{1S}$-trained muon MVA identification with $\Lxy<0.015$\unit{cm} (blue, solid), and the $\PJGy$-trained muon MVA identification with $\Lxy/\sigLxy < 3.5$ (red, dashed). The $\PJGy$-trained and $\PGUP{1S}$-trained selections are used in the mass range 1.1--2.6 and 4.2--7.9\GeV, respectively.} 
\label{fig:muu}
\par}
\end{figure}

At smaller values of $\Mmumu$, the dimuon system is more Lorentz boosted and the muons become more collinear, resulting in larger uncertainties in the reconstructed PV position. The $\Lxy$ distribution becomes broader, whereas the distribution of the vertex displacement significance remains stable.
Therefore, we employ different displacement criteria depending on the mass value.  For events with $\Mmumu<2.6\GeV$ we further require that $\Lxy$ divided by its uncertainty ($\sigLxy$) be less than 3.5, and the $\Lxy$ value is not used in the BDT.  Events with $\Mmumu>4.2\GeV$ are required to have $\Lxy<0.015$\unit{cm} to further suppress nonprompt backgrounds.  Figure~\ref{fig:muu} shows the $\Mmumu$ distribution obtained from the dimuon scouting triggers.  The average efficiency of the $\PGUP{1S}$ ($\PJGy$)-trained muon MVA identification is around 85~(90)\% estimated from the $\PGUP{1S}$ ($\PJGy$) resonance. 
The $\PGUP{1S}$-trained muon MVA identification exhibits better performance for $\Mmumu>4.2\GeV$, as it suppresses more background events, while maintaining high signal efficiency. 
Compared with the simple cut-based selection used in the previous CMS search~\cite{Sirunyan:2019wqq}, the MVA identification provides a higher efficiency while still keeping the misidentification rate low, resulting in improvements to the final limits on the cross section by about 30\%. We refer to this as the inclusive selection.

In the search for a light pseudoscalar particle (the 2HDM+S scenario), the signal is produced via the ggF process, which results in a harder \pt distribution compared to the DY process, due to the enhanced probability for a jet to scatter off of an initial state gluon.  Higher values of muon $\pt$ are thus required to further improve the sensitivity for the pseudoscalar signal. The muons are required to have $\pt>5\GeV$ and the muon pair is required to have $\pt>20 \ (35)\GeV$ for $\Mmumu>4.2\GeV$ ($<2.6\GeV$).  The $\PJGy$-trained muon MVA identification and $\Lxy/\sigLxy < 3.5$ are required for both $\Mmumu<2.6\GeV$ and $\Mmumu>4.2\GeV$. We refer to this as the high-\pt selection.   All selection criteria for both categories are summarized in Table~\ref{tab:sel}.

\begin{table}[h]
  {\centering
    \topcaption{Summary of all selection criteria for an event to enter the analysis in inclusive and high-\pt search categories.}
     \renewcommand{\arraystretch}{1.1}
      \begin{tabular}{lcccc}
        \hline
        Preselection                 &  \multicolumn{4}{c}{$\Lxy<0.2$\unit{cm}, $\abs{\eta^{\mu}}<1.9$, Opposite sign} \\
        \hline
        Signal selection                       & \multicolumn{2}{c}{Inclusive} & \multicolumn{2}{c}{high-\pt}\\[0.5ex]
        Mass                    & $\Mmumu<4\GeV$ & $\Mmumu>4\GeV$ & $\Mmumu<4\GeV$ & $\Mmumu>4\GeV$\\
        ${p}^{\mu}_{\mathrm{T}}$              & \multicolumn{2}{c}{$>$4\GeV}      &   \multicolumn{2}{c}{$>$5\GeV} \\
        Muon                        & J/$\psi$ ID       & $\Upsilon$ ID & J/$\psi$ ID & J/$\psi$ ID\\
        Vertex & $\Lxy/\sigLxy<3.5$
	                             & $\Lxy<0.015$\unit{cm} &  $\Lxy/\sigLxy<3.5$  & $\Lxy/\sigLxy<3.5$ \\
        ${p}^{\mu\mu}_{\mathrm{T}}$             & \NA                & \NA    & $>$35\GeV  & $>$20\GeV \\
        \hline
      \end{tabular}
    \label{tab:sel}
  \par}
\end{table}

\begin{figure}[!htb]
{\centering
\includegraphics[width=0.8\linewidth]{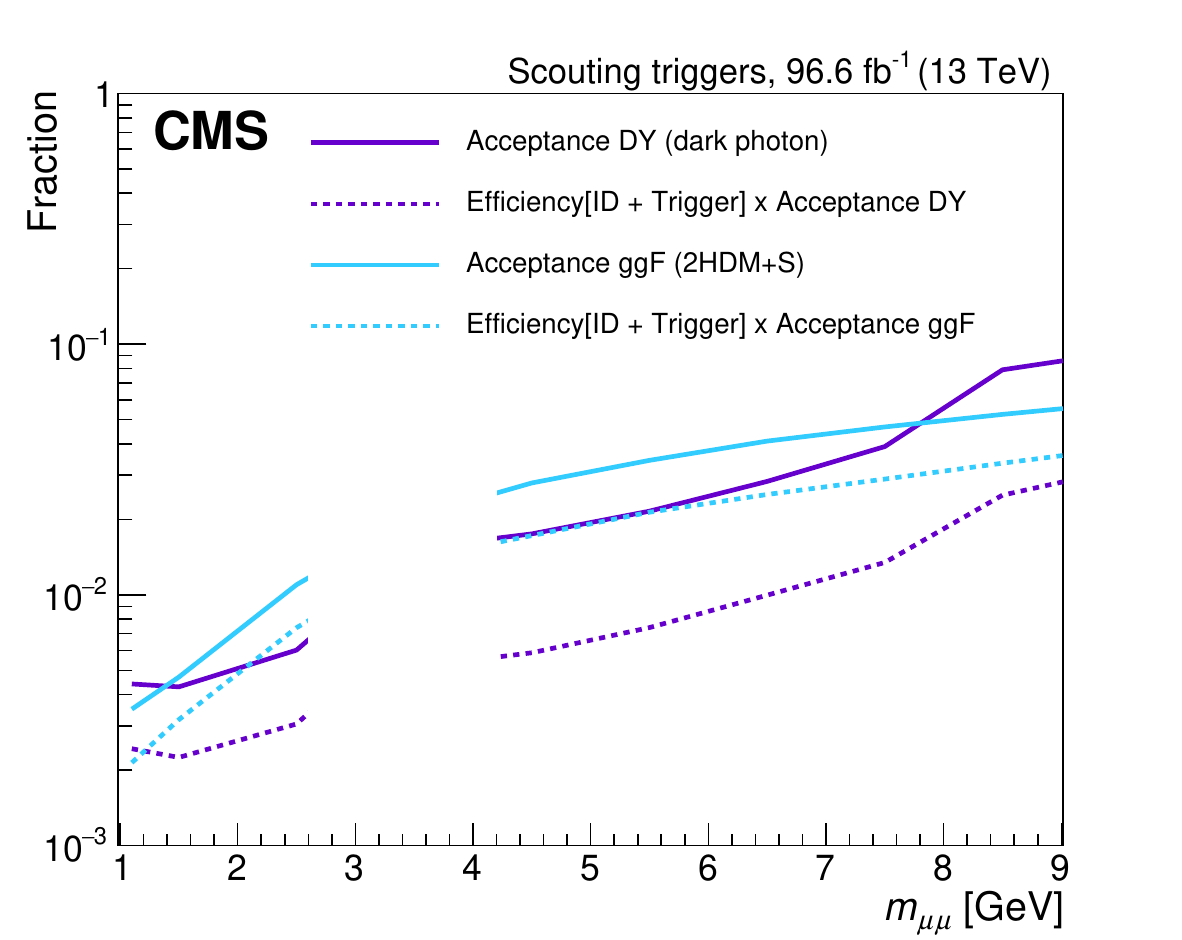}
\caption{The signal acceptance and reconstruction efficiency are extracted from the dark photon (purple, dark) and pseudoscalar (cyan, light) simulations. The region at 2.6--4.2\GeV is excluded because of the presence of the $\PJGy$ and $\Pgy$ resonances.}
\label{fig:acceptance}
\par}
\end{figure}

The muon reconstruction and identification efficiency measured from data and the acceptance extracted from signal simulations are shown in Fig.~\ref{fig:acceptance}.  The efficiency of the vertex displacement requirement of the signal is corrected using the $\PJGy$ and $\PGUP{1S}$ samples from data and simulation.  The total identification efficiency of the event is the product of the identification efficiencies of the two muons.

\section{Signal and background model\label{sec:modelfit}}

The signal is extracted by performing simultaneous signal plus background maximum likelihood fits to the $\Mmumu$ distribution.  For each mass hypothesis, the fit is performed over a mass window spanning $\pm5$ times the mass resolution around the nominal resonance mass. The mass resolution is determined from data to be roughly 1.3\% across the entire mass range.  This window size ensures sufficient sidebands for the background model fit, such that it will not be influenced by the presence of a signal at the central value.  When fitting the mass distributions in the high-$\pt$ selection category, the size of the window is increased to $\pm8$ times the mass resolution, to offset the fact that the number of events in the sideband is significantly reduced by the high $\pt$ requirements.  The probability density function (pdf) chosen to model the signal distribution is the sum of a double Crystal Ball~\cite{DCB, DCB_MO} and a Gaussian distribution.  We limit the search to the case of narrow resonances with an intrinsic width much smaller than the experimental resolution.  Since this same approximation holds for charmonium and $\PGU$ resonances, the shape parameters for the signal pdf are determined from least square mass fits for the $\PJGy$, $\Pgy$, and $\PGU$((1S), (2S), and (3S)) resonances.  The signal pdf parameters, including the mass resolution, are found to have little dependence on the resonance mass or the selection categories.  The only free parameter of the signal pdf in the final fit is the overall normalization, which is proportional to the production cross section of the new particle.

Several possible functions to model the background are investigated.  We use empirical functions because the shape of the background in the search region is complicated by the presence of multiple known resonances and by the acceptance of the various triggers. The families of functions considered are Bernstein polynomials~\cite{ref:bernstein}, a polynomial times an exponential, a sum of exponentials, and a sum of a Bernstein polynomial and a power law function.  The optimal order for each function family is chosen by performing a Fisher $F$-test~\cite{ref:ftest} with background-only fits to the dimuon mass distribution for each mass hypothesis.  The order of the function family is chosen as the lowest one that keeps the number of fits with an $F$ value above the critical value below the false-rejection probability ($\alpha<0.05$). Across all the considered families, the optimal order for the polynomial component is always 4th order or lower. For the sum-of-exponentials function, the optimal number of terms is found to be 7.  The potential bias of each function family is evaluated using simulations assuming either no signal or the presence of a small hypothetical signal.  The bias for each function type, defined as the difference between the assumed and measured signal strength, is less than $\approx$20\% of the statistical uncertainty in the signal strength.  Therefore, the added uncertainty due to the bias of the background function is negligible and is not included in the final fit.  The discrete profiling method~\cite{Dauncey:2014xga} is used to further suppress the bias, which arises from the choice of the background function, by allowing the signal shape to vary between the options listed above during the maximum likelihood fit.  The background model fit of the mass distribution becomes unreliable when the tails of $\PJGy$ and $\Pgy$ resonances enter the mass window, so the mass range 2.6--4.2\GeV is excluded from the search.

Among all the known resonances, the $\PDz$ meson contributes an apparent peak in the $\Mmumu$ continuum in the search range. 
When the pions and kaons from the $\PDz\to \PKp\PKm$ or $\PDz\to \PKm\PGpp$ decays are misreconstructed as muons, the $\Mmumu$ distribution exhibits an enhancement at 1.72 or 1.58\GeV, respectively.  The contamination of these processes in the signal region is estimated from a control region by a simultaneous fit to both regions.  The control region is defined by requiring an $\Lxy/\sigLxy$ to be within 3.5--11.0.  The pdf of the $\PDz\to \PKp\PKm(\PKm\PGpp)$ dimuon peak is derived from simulation.  The normalization is a free parameter that is constrained by a transfer factor between the signal and control regions from the simulation.  The uncertainty in the transfer factor includes the statistical uncertainty of the simulated samples and the mismodeling of $\Lxy/\sigLxy$, which is quantified using the $\PJGy$ peak in both data and simulation.  The $\PJGy$ events in the data are determined by subtracting the contribution from background and nonprompt $\PJGy$ production, estimated through extrapolation from the distribution, where $8<\Lxy/\sigLxy<20$. The contribution from the $\PDz\to \PGpp\PGpm$ process is not modeled because pions are much less likely to be misidentified as muons compared to kaons and there are insufficient $\PDz\to \PGpp\PGpm$ yields in the control region to properly constrain the process.
\section{Systematic uncertainties\label{sec:uncertainty}}

The signal extraction in this analysis is affected by several sources of systematic uncertainties.  The measured integrated luminosity has an uncertainty of 2.3~(2.5)\% for data collected in 2017 (2018)~\cite{CMS-PAS-LUM-17-001,CMS-PAS-LUM-18-002,CMS:LUM-17-003}.  Roughly half of these values are attributed to correlated systematic effects between the years, and the remaining portion is uncorrelated, giving a combined uncertainty of about 2.0\% between the two years.  The uncertainty in the efficiency of the PV displacement requirement is 3\% for $\Mmumu>4.2\GeV$ and is negligible for $\Mmumu<2.6\GeV$. 

The efficiency of the dimuon scouting triggers is measured with respect to fully reconstructed muons using events selected with independent standard triggers. The uncertainties in the trigger efficiency depend on the dimuon mass and vary from 1--16\%, as determined from the difference between the trigger efficiency in data and simulation in each mass bin.

The efficiency of muon identification is measured with respect to trigger requirements. The uncertainties in the identification efficiency of the two muons vary from 4--20\%, and include the statistical uncertainty of the measurement sample and the uncertainty associated with the modeling in the input parameters of the muon MVA identification between the data and simulation.  The uncertainty is largest for muons with low \pt and low $\Delta R$ because of the reduced number of muons in this region in the simulated samples.

The per muon identification efficiency is applied to calculate the total reconstruction efficiency. The identification efficiency is measured from the resonances produced mainly through the gluon splitting process, though the signal production may proceed via other mechanisms. The identification efficiencies between the two muons are correlated because event observables are used in the MVA identification, whereas the efficiencies are treated as uncorrelated when calculating the total efficiency.  To account for the efficiency variation caused by these effects, an additional 4~(8)\% systematic uncertainty is assigned for $\Mmumu>4.2\GeV$~($<2.6\GeV$).  The uncertainty in the transfer factor that connects the normalizations between the signal and control regions of the $\PDz\to \PKp\PKm$($\PKm\PGpp$) backgrounds is 20~(25)\%.

The mass resolution of the signal is measured using different resonances in data. The variation in the observed mass resolutions is around 20\% and is assigned as the systematic uncertainty.  The signal shape uncertainty, which varies from 1.3--2.7\%, is derived by using a double Gaussian as the signal template in the fit.  
As described in Section~\ref{sec:modelfit}, the uncertainty due to the bias of all of the candidate background functions is negligible. The implicit bias associated with having chosen empirical functions to model the background is accounted for via the implementation of the discrete profiling method. This methodology introduces an additional discrete nuisance parameter, associated with the choice of background function, into the maximum likelihood fit.
Table~\ref{tab:sys_summary} summarizes the systematic uncertainties in the search.

\begin{table}[thp]
  {\centering
  \caption{Summary of the experimental systematic uncertainties for a signal model in the model independent search for a dimuon resonance.}
    \begin{tabular}{lcc}
      \hline
      Effect & $\Mmumu<2.6\GeV$ & $\Mmumu>4.2\GeV$\\
      \hline
      Integrated luminosity             &  \multicolumn{2}{c}{2.3--2.5\%} \\
      Mass resolution                   &  \multicolumn{2}{c}{20\%} \\
      Signal shape                      &  1.3--2.7\% & 1.4--1.7\% \\
      Trigger efficiency                &  \multicolumn{2}{c}{1--16\%} \\
      Muon ID efficiency                &   4--9\%    & 12--20\% \\
      Vertex selection                  &       \NA    &      3\% \\
      Efficiency application            &      8\%    &      4\% \\
      \PDz meson normalization          & 20--25\%    &       \NA \\
      \hline
    \end{tabular}
    \label{tab:sys_summary}
  \par}
\end{table}

The theoretical uncertainties in the signal models are not considered in the model-independent search since the propagated kinematic variations exert a negligible influence on the reconstruction efficiencies. Likewise, these uncertainties are not included in the model-dependent limits, to factorize the theoretical and experimental uncertainties. 
\section{Results\label{sec:limit}}

The data are consistent with the continuum dimuon production.  Limits at $95\%$ confidence level (\CL) are set on the product of the signal cross section, the branching fraction to a muon pair, and fiducial acceptance of a hypothetical resonance with a mass from 1.1--2.6\GeV or from 4.2--7.9\GeV.  The computation of exclusion upper limits is performed using the \CLs criterion in the asymptotic approximation~\cite{Junk:1999kv,Read:2002hq,Cowan:2010js}.  A profile likelihood ratio test statistic is built including the systematic uncertainties presented above as nuisance parameters.   A log-normal pdf is used to describe each constrained nuisance parameter.

\begin{figure}[!htb]
{\centering
\includegraphics[width=0.49\linewidth]{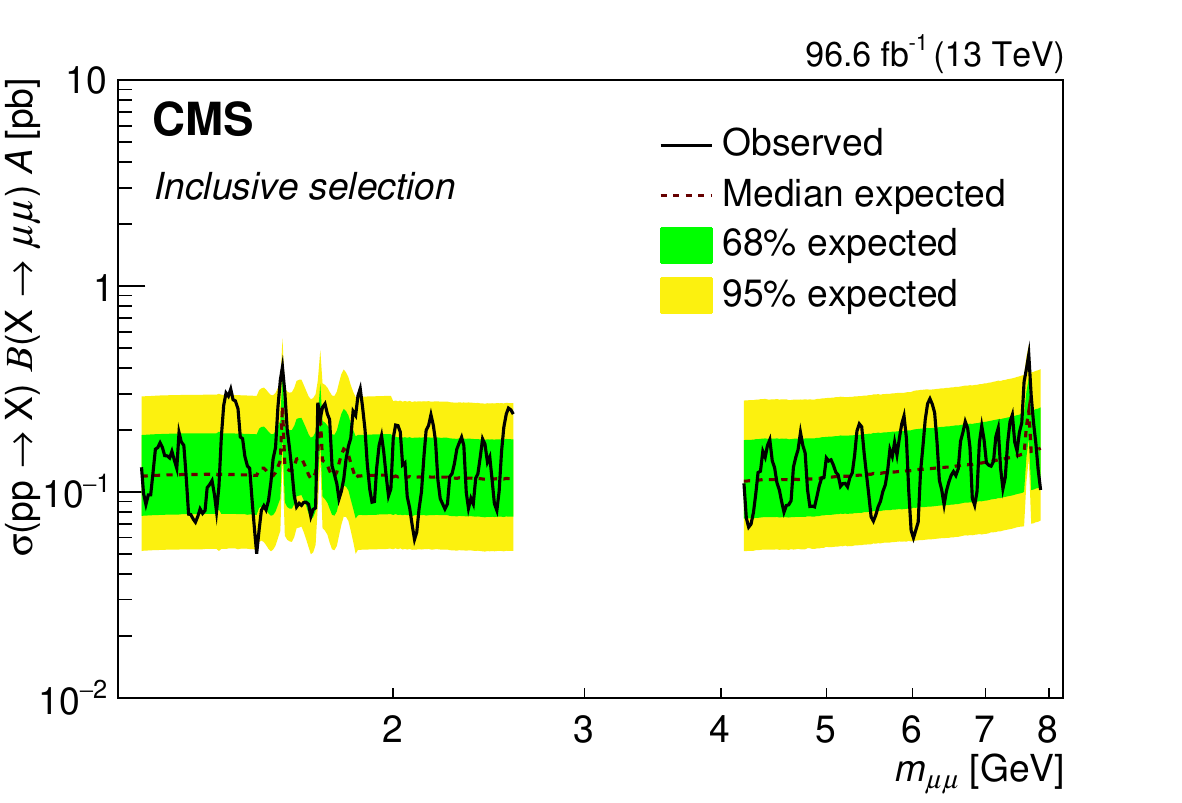}
\includegraphics[width=0.49\linewidth]{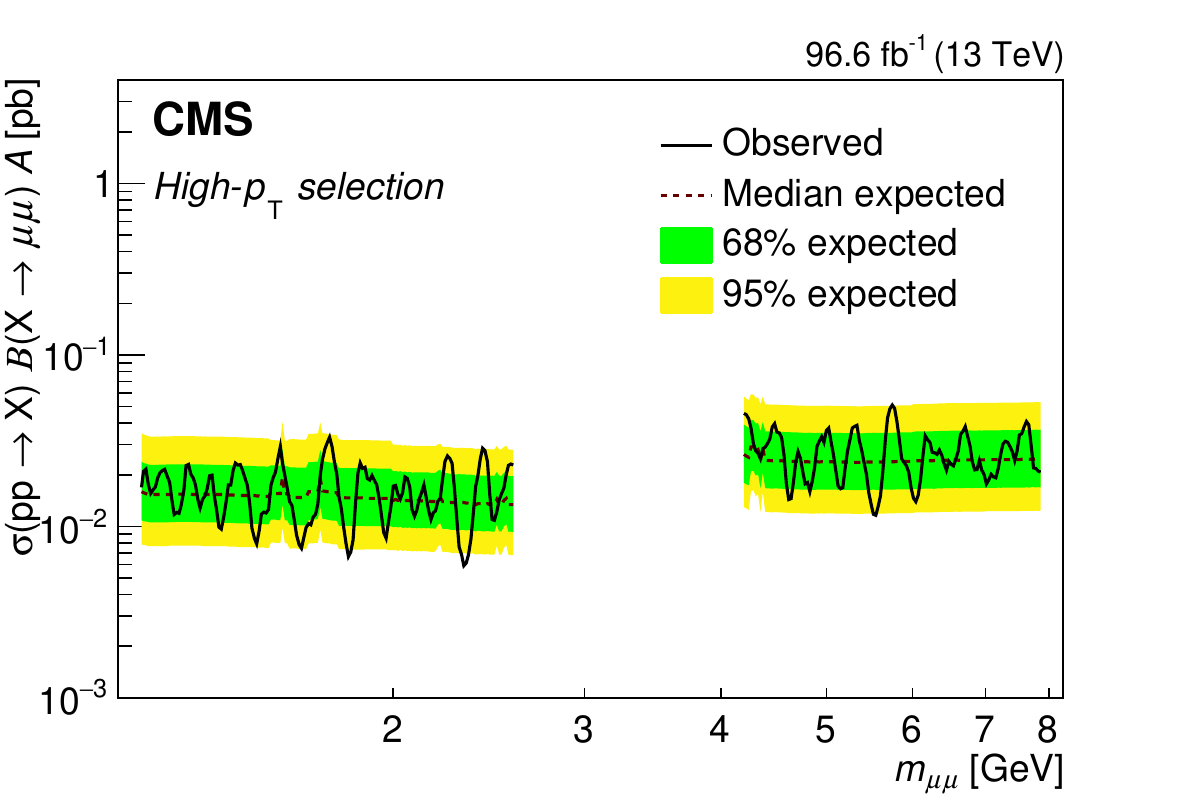}
\caption{Left: Expected and observed model-independent upper limits at 95\% \CL on the product of the signal cross section ($\sigma$), the branching fraction to a pair of muons for the inclusive dimuon selection ($\mathcal{B}$), and fiducial acceptance ($A$). Right: The model-independent limits for the high-\pt selection. The mass region dominated by the $\PJGy$ and $\Pgy$ resonances is excluded from the search.} 
\label{fig:limits}
\par}
\end{figure}

Figure~\ref{fig:limits} shows the expected and observed model independent limits for the inclusive and high-\pt selections.  The 95\% confidence level (\CL) upper limit for the inclusive selection is of order 0.10~(0.15)\unit{pb} in the mass range 1.1--2.6~(4.2--7.9)\GeV.  The 95\% \CL upper limit for the high-\pt selection is about 0.015~(0.024)\unit{pb} in the mass range 1.1--2.6~(4.2--7.9)\GeV. The discontinuities in the expected limits arise from the discrete profiling method applied when different background functions are favored during the fitting process. The upward deviations in the expected limits below 2\GeV are due to the peaking background associated with $\PDz$ meson decays.  The excess at 1.867\GeV in the inclusive selection may correspond to the $\PDz\to \PGpp\PGpm$ or $\PDz\to \PGmp\PGmm$ processes. Distinguishing between these two possibilities is beyond the scope of this paper. 
The observed limits exhibit oscillatory structures within some mass ranges, which result from statistical fluctuations modulated by the sliding mass window. 
No significant excess is observed beyond 2.6 (0.6) standard deviations local (global) significance from the SM background expectation in the high-\pt category.
The model independent limits are exploited to constrain the dark photon~\cite{DIENES1997104, GALISON1984279, HOLDOM1986196} and 2HDM+S~\cite{Haisch:2018kqx} scenarios.

\begin{figure}[!htb]
{\centering
\includegraphics[width=0.95\linewidth]{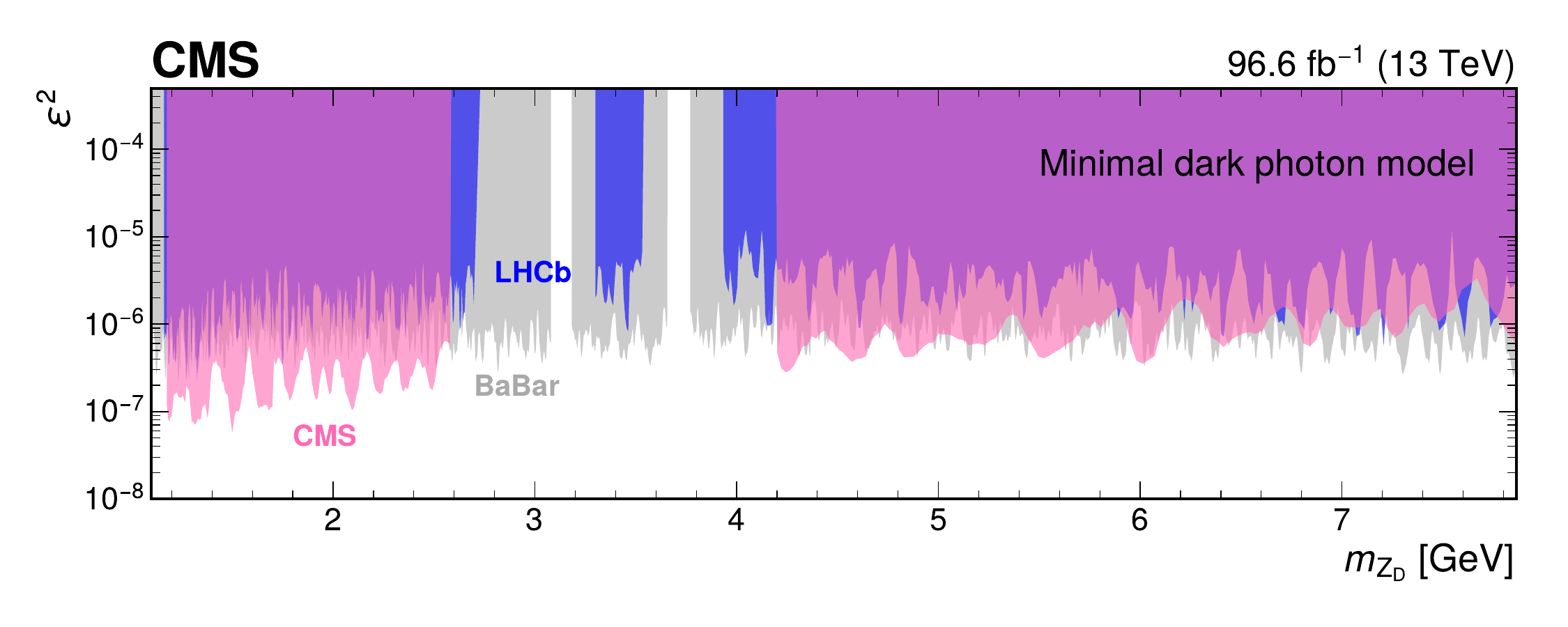}
\caption{Observed upper limits at 90\% \CL on the square of the kinetic mixing coefficient $\varepsilon$ in the minimal model of a dark photon from the CMS search in the mass ranges of 1.1--2.6\GeV and 4.2--7.9\GeV (pink). The CMS limits are compared with the existing limits at 90\% \CL provided by LHCb~\cite{Aaij:2019bvg} (blue) and BaBar~\cite{Lees:2014xha} (gray). The LHCb exclusion limit also extends below 1\GeV and to smaller couplings through its search for long-lived signals.}
\label{fig:darkphoton_limits}
\par}
\end{figure}

In the first case, limits on the squared kinetic mixing coefficient \epsSq are extracted as a function of the dark photon mass ($m_{\Zdark}$) according to the following relation:
\begin{linenomath}
\begin{equation}
\label{eq:epsilon_limit}
 \begin{aligned}
  \sigma_{\Pp\Pp\to\Zdark}  \varepsilon^2   \mathcal{B}  A = \sigma_{\mathrm{limit}},
 \end{aligned}                         
\end{equation}
\end{linenomath}
where $\sigma_{\Pp\Pp\to\Zdark}$ is the dark photon cross section with $\varepsilon=1$, $\mathcal{B}$ is the branching fraction of $\Zdark\to\PGmp\PGmm$, $\sigma_{\mathrm{limit}}$ is the model independent limit, and $A$ is the acceptance.  We use the numerical software \DYTurbo~\cite{dyturbo} to generate the DY process at next-to-next-to-LO (NNLO) with the renormalization and factorization scales set to $\mu=0.5\sqrt{\smash[b]{\Mmumu^2+{\pt}_{\PGm\PGm}^2}}$ and to correct the dark photon cross section and extract the acceptance, because the dark photon is produced through the DY-like process.  The interpretation of the cross section limit as a limit on the mixing coefficient \epsSq is subject to theoretical uncertainties.  Uncertainties in the theoretical cross section are found by varying the renormalization and factorization scales up and down by a factor of 2, and the uncertainty in the acceptance is found by comparing the results between the \DYTurbo and \MGvATNLO generators.  The scale variations are between 3.5 and 9.7\%, whereas the difference in acceptance is up to 31\%.  Upper limits at $90\%$ \CL on the mixing coefficient \epsSq are presented in Fig.~\ref{fig:darkphoton_limits}, together with the LHCb and BaBar limits reinterpreted by \textsc{DarkCast}~\cite{darkcast}. The sensitivity becomes better at low masses because of the larger DY production cross section at lower energy scale.  The kinetic mixing coefficient $\varepsilon$ reflects the degree of mixing and the strength of the coupling of dark photon to other SM particles.  The mixing coefficient \epsSq above $2 \ (9)\times 10^{-7}$ is excluded at $m_{\Zdark}=2 \ (7)\GeV$.  These results complement the previous CMS search at larger masses~\cite{Sirunyan:2019wqq}.
                                 
In the 2HDM+S scenarios, the limits from the high-\pt selection are converted to an upper limit on the mixing angle $\thetaH$ as a function of the light pseudoscalar-boson mass. The recasting of the limit is performed under the assumption of $\tan\beta=0.5$ according to the following relation:
\begin{linenomath}
\begin{equation}
\label{eq:sentheta_limit}
 \begin{aligned}
  \sigma_{\Pp\Pp\to \mathrm{a}}  \sin^{2}(\thetaH) \mathcal{B} A = \sigma_{\mathrm{limit}},
 \end{aligned}
\end{equation}
\end{linenomath}
where $\sigma_{\Pp\Pp\to \mathrm{a}}$ is the production cross section of the light pseudoscalar \Pa, $\sigma_{\mathrm{limit}}$ is the model independent limit, $\mathcal{B}$ is the branching fraction of $\mathrm{a}\to\PGmp\PGmm$, and $A$ is the acceptance.  The $\sigma_{\Pp\Pp\to \mathrm{a}}$ is computed from the \HIGLU generator at NNLO using the NNPDF3.0 PDF set~\cite{nnpdf} and assuming renormalization and factorization scales $\mu=0.5 m_\mathrm{a}$, as in the LHCb search~\cite{Aaij:2020ikh}. Uncertainties in the theoretical cross section for pseudoscalar production via ggF are found by varying the renormalization and factorization scales by a factor of 2.  These uncertainties are around 90\% at $m_\mathrm{a}=1.18\GeV$ and gradually reduce to 10\% at $m_\mathrm{a}>4.2\GeV$.  The uncertainty in the ggF acceptance is estimated by comparing the values obtained with \MGvATNLO and \PYTHIA, and is about 30\%. 

\begin{figure}[!htb]
{\centering
\includegraphics[width=0.95\linewidth]{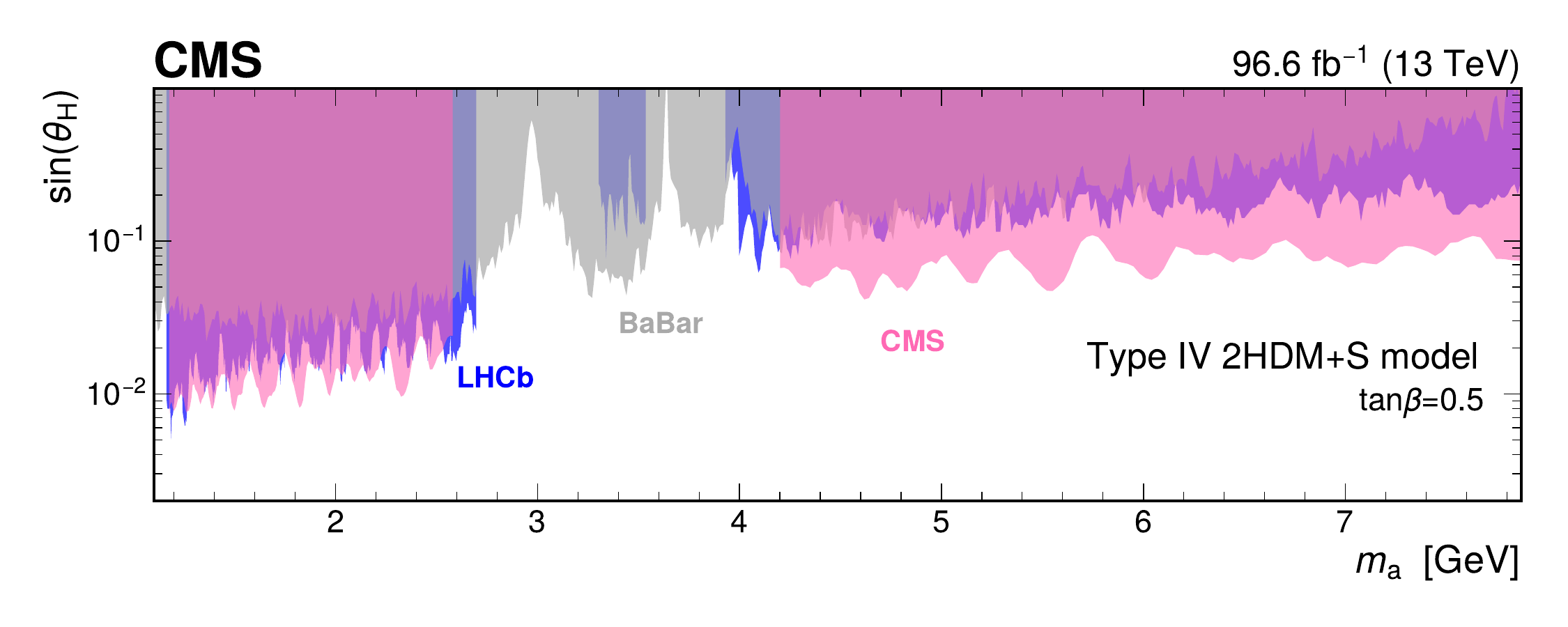}
\caption{Observed upper limits at 90\% \CL on the mixing angle $\thetaH$ for the 2HDM+S scenario from the CMS search in the mass ranges of 1.1--2.6\GeV and 4.2--7.9\GeV (pink). The CMS limits are compared with the existing limits at 90\% \CL provided by LHCb~\cite{Aaij:2020ikh} (blue) and BaBar~\cite{Lees:2014xha} (gray).}
\label{fig:sentheta_limit}
\par}
\end{figure}

Observed upper limits at $90\%$ \CL are presented in Fig.~\ref{fig:sentheta_limit}.  Values of $\sin(\thetaH)$ above $\approx$0.01 (0.02) are excluded at $m_{\mathrm{a}}=2 \ (7)\GeV$ with fixed $\tan\beta=0.5$. The limits derived from this search in the low-mass region are competitive with recently reported results from the LHCb experiment~\cite{Aaij:2020ikh} below the charmonium peaks and better above them.

\section{Summary\label{sec:summary}}

A search for direct production of a narrow resonance decaying to a pair of muons has been presented using proton-proton collision data recorded by the CMS experiment at $\sqrt{s}=13\TeV$ in 2017--2018.  The search is performed in the dimuon mass regions of 1.1--2.6\GeV and 4.2--7.9\GeV using data collected in a dedicated high-rate trigger stream, corresponding to an integrated luminosity of 96.6\fbinv.  A multivariate analysis method is used to identify muons to achieve a higher sensitivity.  No significant excess of events above the expectation from the standard model background is observed.  Model-independent limits on production rates of dimuon resonances within the experimental fiducial acceptance are set. 
Competitive or world's best limits are set at 90\% confidence level for a minimal dark photon model and for a scenario with two Higgs doublets and an extra complex scalar singlet (2HDM+S).
Values of the squared kinetic mixing coefficient \epsSq in the dark photon model above $10^{-6}$ are excluded over most of the mass range of the search.  In the 2HDM+S, values of the mixing angle $\sin(\thetaH)$ above 0.08 are excluded over most of the mass range of the search with a fixed ratio of the Higgs doublets vacuum expectation $\tan\beta=0.5$.

\begin{acknowledgments}
We thank Stefano Camarda for suggesting the \DYTurbo settings for calculating the DY cross section at low mass, for the dark photon interpretation. We also thank Philip Ilten for providing the details of the $\Pp\Pp\to\Pa$ cross section calculation using the \HIGLU generator, for the comparison of CMS and LHCb results.

We congratulate our colleagues in the CERN accelerator departments for the excellent performance of the LHC and thank the technical and administrative staffs at CERN and at other CMS institutes for their contributions to the success of the CMS effort. In addition, we gratefully acknowledge the computing centers and personnel of the Worldwide LHC Computing Grid and other centers for delivering so effectively the computing infrastructure essential to our analyses. Finally, we acknowledge the enduring support for the construction and operation of the LHC, the CMS detector, and the supporting computing infrastructure provided by the following funding agencies: SC (Armenia), BMBWF and FWF (Austria); FNRS and FWO (Belgium); CNPq, CAPES, FAPERJ, FAPERGS, and FAPESP (Brazil); MES and BNSF (Bulgaria); CERN; CAS, MoST, and NSFC (China); MINCIENCIAS (Colombia); MSES and CSF (Croatia); RIF (Cyprus); SENESCYT (Ecuador); MoER, ERC PUT and ERDF (Estonia); Academy of Finland, MEC, and HIP (Finland); CEA and CNRS/IN2P3 (France); SRNSF (Georgia); BMBF, DFG, and HGF (Germany); GSRI (Greece); NKFIH (Hungary); DAE and DST (India); IPM (Iran); SFI (Ireland); INFN (Italy); MSIP and NRF (Republic of Korea); MES (Latvia); LAS (Lithuania); MOE and UM (Malaysia); BUAP, CINVESTAV, CONACYT, LNS, SEP, and UASLP-FAI (Mexico); MOS (Montenegro); MBIE (New Zealand); PAEC (Pakistan); MES and NSC (Poland); FCT (Portugal); MESTD (Serbia); MCIN/AEI and PCTI (Spain); MOSTR (Sri Lanka); Swiss Funding Agencies (Switzerland); MST (Taipei); MHESI and NSTDA (Thailand); TUBITAK and TENMAK (Turkey); NASU (Ukraine); STFC (United Kingdom); DOE and NSF (USA).

\hyphenation{Rachada-pisek} Individuals have received support from the Marie-Curie program and the European Research Council and Horizon 2020 Grant, contract Nos.\ 675440, 724704, 752730, 758316, 765710, 824093, and COST Action CA16108 (European Union); the Leventis Foundation; the Alfred P.\ Sloan Foundation; the Alexander von Humboldt Foundation; the Science Committee, project no. 22rl-037 (Armenia); the Belgian Federal Science Policy Office; the Fonds pour la Formation \`a la Recherche dans l'Industrie et dans l'Agriculture (FRIA-Belgium); the Agentschap voor Innovatie door Wetenschap en Technologie (IWT-Belgium); the F.R.S.-FNRS and FWO (Belgium) under the ``Excellence of Science -- EOS" -- be.h project n.\ 30820817; the Beijing Municipal Science \& Technology Commission, No. Z191100007219010 and Fundamental Research Funds for the Central Universities (China); the Ministry of Education, Youth and Sports (MEYS) of the Czech Republic; the Shota Rustaveli National Science Foundation, grant FR-22-985 (Georgia); the Deutsche Forschungsgemeinschaft (DFG), under Germany's Excellence Strategy -- EXC 2121 ``Quantum Universe" -- 390833306, and under project number 400140256 - GRK2497; the Hellenic Foundation for Research and Innovation (HFRI), Project Number 2288 (Greece); the Hungarian Academy of Sciences, the New National Excellence Program - \'UNKP, the NKFIH research grants K 124845, K 124850, K 128713, K 128786, K 129058, K 131991, K 133046, K 138136, K 143460, K 143477, 2020-2.2.1-ED-2021-00181, and TKP2021-NKTA-64 (Hungary); the Council of Science and Industrial Research, India; the Latvian Council of Science; the Ministry of Education and Science, project no. 2022/WK/14, and the National Science Center, contracts Opus 2021/41/B/ST2/01369 and 2021/43/B/ST2/01552 (Poland); the Funda\c{c}\~ao para a Ci\^encia e a Tecnologia, grant CEECIND/01334/2018 (Portugal); the National Priorities Research Program by Qatar National Research Fund; MCIN/AEI/10.13039/501100011033, ERDF ``a way of making Europe", and the Programa Estatal de Fomento de la Investigaci{\'o}n Cient{\'i}fica y T{\'e}cnica de Excelencia Mar\'{\i}a de Maeztu, grant MDM-2017-0765 and Programa Severo Ochoa del Principado de Asturias (Spain); the Chulalongkorn Academic into Its 2nd Century Project Advancement Project, and the National Science, Research and Innovation Fund via the Program Management Unit for Human Resources \& Institutional Development, Research and Innovation, grant B05F650021 (Thailand); the Kavli Foundation; the Nvidia Corporation; the SuperMicro Corporation; the Welch Foundation, contract C-1845; and the Weston Havens Foundation (USA).
\end{acknowledgments}

\bibliography{auto_generated}

\providecommand{\href}[2]{#2}\begingroup\raggedright\begin{thebibliography}{10}%
\makeatletter
\providecommand{\hrefCMSnoop }[0]{\@secondoftwo}%
\makeatother
\providecommand{\doi}{\texttt{doi:}\begingroup \urlstyle{tt}\Url}

\bibitem{PhysRevLett.33.1404}
J.~J. Aubert\hrefCMSnoop {}{ {et~al.}, ``{Experimental observation of a heavy
  particle $J$}'',} \textit{ Phys. Rev. Lett.} \textbf{ 33} (1974) 1404,
  \href{http://dx.doi.org/10.1103/PhysRevLett.33.1404}{\doi{10.1103/PhysRevLett.33.1404}}.

\bibitem{PhysRevLett.33.1406}
\hrefCMSnoop {}{J.~E. Augustin {et~al.}, ``{Discovery of a narrow resonance in
  $\text{e}^{+}\text{e}^{\ensuremath{-}}$ annihilation}'',} \textit{ Phys. Rev.
  Lett.} \textbf{ 33} (1974) 1406,
  \href{http://dx.doi.org/10.1103/PhysRevLett.33.1406}{\doi{10.1103/PhysRevLett.33.1406}}.

\bibitem{PhysRevLett.39.252}
\hrefCMSnoop {}{S.~W. Herb {et~al.}, ``{Observation of a dimuon resonance at
  9.5 GeV in 400-GeV proton-nucleus collisions}'',} \textit{ Phys. Rev. Lett.}
  \textbf{ 39} (1977) 252,
  \href{http://dx.doi.org/10.1103/PhysRevLett.39.252}{\doi{10.1103/PhysRevLett.39.252}}.

\bibitem{1983398}
\hrefCMSnoop {}{{UA1} Collaboration, ``{Experimental observation of lepton
  pairs of invariant mass around 95 GeV/c$^2$ at the CERN SPS collider}'',}
  \textit{ Phys. Lett. B} \textbf{ 126} (1983) 398,
  \href{http://dx.doi.org/10.1016/0370-2693(83)90188-0}{\doi{10.1016/0370-2693(83)90188-0}}.

\bibitem{Aad:2012tfa}
\hrefCMSnoop {}{{ATLAS Collaboration}, ``{Observation of a new particle in the
  search for the Standard Model Higgs boson with the ATLAS detector at the
  LHC}'',} \textit{ Phys. Lett. B} \textbf{ 716} (2012) 1,
  \href{http://dx.doi.org/10.1016/j.physletb.2012.08.020}{\doi{10.1016/j.physletb.2012.08.020}},
  \href{http://www.arXiv.org/abs/1207.7214}{\texttt{arXiv:1207.7214}}.

\bibitem{Chatrchyan:2012ufa}
\hrefCMSnoop {}{{CMS Collaboration}, ``{Observation of a new boson at a mass of
  125 GeV with the CMS experiment at the LHC}'',} \textit{ Phys. Lett. B}
  \textbf{ 716} (2012) 30,
  \href{http://dx.doi.org/10.1016/j.physletb.2012.08.021}{\doi{10.1016/j.physletb.2012.08.021}},
  \href{http://www.arXiv.org/abs/1207.7235}{\texttt{arXiv:1207.7235}}.

\bibitem{Chatrchyan:2013}
\hrefCMSnoop {}{{CMS Collaboration}, ``{Observation of a new boson with mass
  near 125 GeV in pp collisions at $ \sqrt{s}=7 $and 8 TeV}'',} \textit{ JHEP}
  \textbf{ 06} (2013) 81,
  \href{http://dx.doi.org/10.1007/JHEP06(2013)081}{\doi{10.1007/JHEP06(2013)081}},
  \href{http://www.arXiv.org/abs/1303.4571}{\texttt{arXiv:1303.4571}}.

\bibitem{Aad:2019fac}
\hrefCMSnoop {}{{ATLAS Collaboration}, ``{Search for high-mass dilepton
  resonances using 139 fb$^{-1}$ of pp collision data collected at $\sqrt{s} =
  13$ TeV with the ATLAS detector}'',} \textit{ Phys. Lett. B} \textbf{ 796}
  (2019) 68,
  \href{http://dx.doi.org/10.1016/j.physletb.2019.07.016}{\doi{10.1016/j.physletb.2019.07.016}},
  \href{http://www.arXiv.org/abs/1903.06248}{\texttt{arXiv:1903.06248}}.

\bibitem{Sirunyan:2021khd}
\hrefCMSnoop {}{{CMS Collaboration}, ``{Search for resonant and nonresonant new
  phenomena in high-mass dilepton final states at $\sqrt{s} = $ 13 TeV}'',}
  \textit{ JHEP} \textbf{ 07} (2021) 208,
  \href{http://dx.doi.org/10.1007/JHEP07(2021)208}{\doi{10.1007/JHEP07(2021)208}},
  \href{http://www.arXiv.org/abs/2103.02708}{\texttt{arXiv:2103.02708}}.

\bibitem{ArkaniHamed:2008qn}
\hrefCMSnoop {}{N.~Arkani-Hamed, D.~P. Finkbeiner, T.~R. Slatyer, and
  N.~Weiner, ``{A theory of dark matter}'',} \textit{ Phys. Rev. D} \textbf{
  79} (2009) 015014,
  \href{http://dx.doi.org/10.1103/PhysRevD.79.015014}{\doi{10.1103/PhysRevD.79.015014}},
  \href{http://www.arXiv.org/abs/0810.0713}{\texttt{arXiv:0810.0713}}.

\bibitem{Lees:2014xha}
\hrefCMSnoop {}{{BaBar} Collaboration, ``{Search for a dark photon in
  $\text{e}^+\text{e}^-$ collisions at BaBar}'',} \textit{ Phys. Rev. Lett.}
  \textbf{ 113} (2014) 201801,
  \href{http://dx.doi.org/10.1103/PhysRevLett.113.201801}{\doi{10.1103/PhysRevLett.113.201801}},
  \href{http://www.arXiv.org/abs/1406.2980}{\texttt{arXiv:1406.2980}}.

\bibitem{Aaij:2017rft}
\hrefCMSnoop {}{{LHCb Collaboration}, ``{Search for dark photons produced in 13
  TeV pp collisions}'',} \textit{ Phys. Rev. Lett.} \textbf{ 120} (2018)
  061801,
  \href{http://dx.doi.org/10.1103/PhysRevLett.120.061801}{\doi{10.1103/PhysRevLett.120.061801}},
  \href{http://www.arXiv.org/abs/1710.02867}{\texttt{arXiv:1710.02867}}.

\bibitem{Aaij:2019bvg}
\hrefCMSnoop {}{{LHCb Collaboration}, ``{Search for $A'\to\mu^+\mu^-$
  decays}'',} \textit{ Phys. Rev. Lett.} \textbf{ 124} (2020) 041801,
  \href{http://dx.doi.org/10.1103/PhysRevLett.124.041801}{\doi{10.1103/PhysRevLett.124.041801}},
  \href{http://www.arXiv.org/abs/1910.06926}{\texttt{arXiv:1910.06926}}.

\bibitem{Aaij:2020ikh}
\hrefCMSnoop {}{{LHCb Collaboration}, ``{Searches for low-mass dimuon
  resonances}'',} \textit{ JHEP} \textbf{ 10} (2020) 156,
  \href{http://dx.doi.org/10.1007/JHEP10(2020)156}{\doi{10.1007/JHEP10(2020)156}},
  \href{http://www.arXiv.org/abs/2007.03923}{\texttt{arXiv:2007.03923}}.

\bibitem{Sirunyan:2019wqq}
\hrefCMSnoop {}{{CMS Collaboration}, ``{Search for a narrow resonance lighter
  than 200 GeV decaying to a pair of muons in proton-proton collisions at
  $\sqrt{s} = 13$ TeV}'',} \textit{ Phys. Rev. Lett.} \textbf{ 124} (2020)
  131802,
  \href{http://dx.doi.org/10.1103/PhysRevLett.124.131802}{\doi{10.1103/PhysRevLett.124.131802}},
  \href{http://www.arXiv.org/abs/1912.04776}{\texttt{arXiv:1912.04776}}.

\bibitem{CMS:scouting}
\hrefCMSnoop {}{{CMS Collaboration}, ``{Search for narrow resonances in dijet
  final states at $\sqrt{s} = 8$ TeV with the novel CMS technique of data
  scouting}'',} \textit{ Phys. Rev. Lett.} \textbf{ 117} (2016) 031802,
  \href{http://dx.doi.org/10.1103/PhysRevLett.117.031802}{\doi{10.1103/PhysRevLett.117.031802}},
  \href{http://www.arXiv.org/abs/1604.08907}{\texttt{arXiv:1604.08907}}.

\bibitem{GALISON1984279}
\hrefCMSnoop {}{P.~Galison and A.~Manohar, ``{Two Z's or not two Z's?}'',}
  \textit{ {Phys. Lett. B}} \textbf{ 136} (1984) 279,
  \href{http://dx.doi.org/10.1016/0370-2693(84)91161-4}{\doi{10.1016/0370-2693(84)91161-4}}.

\bibitem{HOLDOM1986196}
\hrefCMSnoop {}{B.~Holdom, ``{Two U(1)'s and $\epsilon$ charge shifts}'',}
  \textit{ {Phys. Lett. B}} \textbf{ 166} (1986) 196,
  \href{http://dx.doi.org/10.1016/0370-2693(86)91377-8}{\doi{10.1016/0370-2693(86)91377-8}}.

\bibitem{hahm1}
\hrefCMSnoop {}{D.~Curtin, R.~Essig, S.~Gori, and J.~Shelton, ``{Illuminating
  dark photons with high-energy colliders}'',} \textit{ JHEP} \textbf{ 02}
  (2015) 157,
  \href{http://dx.doi.org/10.1007/JHEP02(2015)157}{\doi{10.1007/JHEP02(2015)157}},
  \href{http://www.arXiv.org/abs/1412.0018}{\texttt{arXiv:1412.0018}}.

\bibitem{Haisch:2018kqx}
\hrefCMSnoop {}{U.~Haisch, J.~F. Kamenik, A.~Malinauskas, and M.~Spira,
  ``Collider constraints on light pseudoscalars'',} \textit{ JHEP} \textbf{ 03}
  (2018) 178,
  \href{http://dx.doi.org/10.1007/JHEP03(2018)178}{\doi{10.1007/JHEP03(2018)178}},
  \href{http://www.arXiv.org/abs/1802.02156}{\texttt{arXiv:1802.02156}}.

\bibitem{Argyropoulos:2022ezr}
\hrefCMSnoop {}{S.~Argyropoulos and U.~Haisch, ``{Benchmarking LHC searches for
  light 2HDM+a pseudoscalars}'',} \textit{ SciPost Phys.} \textbf{ 13} (2022)
  007,
  \href{http://dx.doi.org/10.21468/SciPostPhys.13.1.007}{\doi{10.21468/SciPostPhys.13.1.007}},
  \href{http://www.arXiv.org/abs/2202.12631}{\texttt{arXiv:2202.12631}}.

\bibitem{hepdata}
\hrefCMSnoop {}{}{HEPD}ata record for this analysis, 2023.
\newblock
  \href{http://dx.doi.org/10.17182/hepdata.140424}{\doi{10.17182/hepdata.140424}}.

\bibitem{Chatrchyan:2008zzk}
\hrefCMSnoop {}{{CMS Collaboration}, ``The {CMS} experiment at the {CERN}
  {LHC}'',} \textit{ JINST} \textbf{ 3} (2008) S08004,
  \href{http://dx.doi.org/10.1088/1748-0221/3/08/S08004}{\doi{10.1088/1748-0221/3/08/S08004}}.

\bibitem{CMS_trigger}
\hrefCMSnoop {}{{CMS Collaboration}, ``{The CMS trigger system}'',} \textit{
  JINST} \textbf{ 12} (2017) P01020,
  \href{http://dx.doi.org/10.1088/1748-0221/12/01/P01020}{\doi{10.1088/1748-0221/12/01/P01020}},
  \href{http://www.arXiv.org/abs/1609.02366}{\texttt{arXiv:1609.02366}}.

\bibitem{ref:hadronicScouting}
\hrefCMSnoop {}{{CMS Collaboration}, ``{Search for dijet resonances in
  proton-proton collisions at $\sqrt{s} = 13$ TeV and constraints on dark
  matter and other models}'',} \textit{ Phys. Lett. B} \textbf{ 769} (2017)
  520,
  \href{http://dx.doi.org/10.1016/j.physletb.2017.02.012}{\doi{10.1016/j.physletb.2017.02.012}},
  \href{http://www.arXiv.org/abs/1611.03568}{\texttt{arXiv:1611.03568}}.

\bibitem{Mukherjee:2017wcl}
\hrefCMSnoop {}{{CMS Collaboration}, ``{Data scouting: a new trigger
  paradigm}'',} in \textit{ {5th Large Hadron Collider Physics Conference}}.
\newblock 2017.
\newblock
  \href{http://www.arXiv.org/abs/1708.06925}{\texttt{arXiv:1708.06925}}.

\bibitem{Phase1Pixel}
\hrefCMSnoop {}{{CMS Tracker Group}, ``The {CMS} {Phase-1} pixel detector
  upgrade'',} \textit{ JINST} \textbf{ 16} (2021) P02027,
  \href{http://dx.doi.org/10.1088/1748-0221/16/02/P02027}{\doi{10.1088/1748-0221/16/02/P02027}},
  \href{http://www.arXiv.org/abs/2012.14304}{\texttt{arXiv:2012.14304}}.

\bibitem{pythia}
T.~Sj{\"o}strand\hrefCMSnoop {}{ {et~al.}, ``{An introduction to PYTHIA
  8.2}'',} \textit{ Comput. Phys. Commun.} \textbf{ 191} (2015) 159,
  \href{http://dx.doi.org/10.1016/j.cpc.2015.01.024}{\doi{10.1016/j.cpc.2015.01.024}},
  \href{http://www.arXiv.org/abs/1410.3012}{\texttt{arXiv:1410.3012}}.

\bibitem{hahm2}
\hrefCMSnoop {}{S.~Gopalakrishna, S.~Jung, and J.~D. Wells, ``{Higgs boson
  decays to four fermions through an abelian hidden sector}'',} \textit{ Phys.
  Rev. D} \textbf{ 78} (2008) 055002,
  \href{http://dx.doi.org/10.1103/PhysRevD.78.055002}{\doi{10.1103/PhysRevD.78.055002}},
  \href{http://www.arXiv.org/abs/0801.3456}{\texttt{arXiv:0801.3456}}.

\bibitem{hahm3}
D.~Curtin\hrefCMSnoop {}{ {et~al.}, ``{Exotic decays of the 125 GeV Higgs
  boson}'',} \textit{ Phys. Rev. D} \textbf{ 90} (2014) 075004,
  \href{http://dx.doi.org/10.1103/PhysRevD.90.075004}{\doi{10.1103/PhysRevD.90.075004}},
  \href{http://www.arXiv.org/abs/1312.4992}{\texttt{arXiv:1312.4992}}.

\bibitem{madgraph}
J.~Alwall\hrefCMSnoop {}{ {et~al.}, ``The automated computation of tree-level
  and next-to-leading order differential cross sections, and their matching to
  parton shower simulations'',} \textit{ JHEP} \textbf{ 07} (2014) 079,
  \href{http://dx.doi.org/10.1007/JHEP07(2014)079}{\doi{10.1007/JHEP07(2014)079}},
  \href{http://www.arXiv.org/abs/1405.0301}{\texttt{arXiv:1405.0301}}.

\bibitem{nnpdf}
\hrefCMSnoop {}{{NNPDF} Collaboration, ``{Parton distributions for the LHC Run
  II}'',} \textit{ JHEP} \textbf{ 04} (2015) 040,
  \href{http://dx.doi.org/10.1007/JHEP04(2015)040}{\doi{10.1007/JHEP04(2015)040}},
  \href{http://www.arXiv.org/abs/1410.8849}{\texttt{arXiv:1410.8849}}.

\bibitem{cp5bib}
\hrefCMSnoop {}{{CMS Collaboration}, ``{Extraction and validation of a new set
  of CMS PYTHIA8 tunes from underlying-event measurements}'',} \textit{ Eur.
  Phys. J. C} \textbf{ 80} (2020) 4,
  \href{http://dx.doi.org/10.1140/epjc/s10052-019-7499-4}{\doi{10.1140/epjc/s10052-019-7499-4}},
  \href{http://www.arXiv.org/abs/1903.12179}{\texttt{arXiv:1903.12179}}.

\bibitem{geant4}
\hrefCMSnoop {}{{\GEANTfour} Collaboration, ``{$\GEANTfour$ --- a simulation
  toolkit}'',} \textit{ Nucl. Instrum. Meth. A} \textbf{ 506} (2003) 250,
\href{http://dx.doi.org/10.1016/S0168-9002(03)01368-8}{\doi{10.1016/S0168-9002(03)01368-8}}.

\bibitem{bdt}
B.~P. Roe\hrefCMSnoop {}{ {et~al.}, ``{Boosted decision trees, an alternative
  to artificial neural networks}'',} \textit{ Nucl. Instrum. Meth. A} \textbf{
  543} (2005) 577,
  \href{http://dx.doi.org/10.1016/j.nima.2004.12.018}{\doi{10.1016/j.nima.2004.12.018}},
  \href{http://www.arXiv.org/abs/physics/0408124}{\texttt{arXiv:physics/0408124}}.

\bibitem{ref:TnP}
\hrefCMSnoop {}{{CMS Collaboration}, ``{Performance of CMS muon reconstruction
  in pp collision events at $\sqrt{s} = 7$ TeV}'',} \textit{ JINST} \textbf{ 7}
  (2012) P10002,
  \href{http://dx.doi.org/10.1088/1748-0221/7/10/P10002}{\doi{10.1088/1748-0221/7/10/P10002}},
  \href{http://www.arXiv.org/abs/1206.4071}{\texttt{arXiv:1206.4071}}.

\bibitem{DCB}
\href
  {https://slac.stanford.edu/pubs/slacreports/reports02/slac-r-255.pdf}{J.~E.
  Gaiser, ``Charmonium spectroscopy from radiative decays of the J/$\Psi$ and
  $\Psi^{\prime}$''}.
\newblock PhD thesis, Stanford University, 1982.
\newblock {SLAC} Report {SLAC-R-255}.

\bibitem{DCB_MO}
\href {https://www.slac.stanford.edu/cgi-bin/getdoc/slac-r-236.pdf}{M.~J.
  Oreglia, ``A study of the reactions $\Psi^{\prime} \rightarrow
  \gamma\gamma\Psi$''}.
\newblock PhD thesis, Stanford University, 1980.
\newblock {SLAC} Report {SLAC-R-236}.

\bibitem{ref:bernstein}
\hrefCMSnoop {}{S.~Bernstein, ``D\'{e}monstration du th\'{e}or\`{e}me de
  {W}eierstrass fond\'{e}e sur le calcul des probabiliti\'{e}s'',} \textit{
  Comm. Soc. Math. Kharkov} \textbf{ 13} (1912) 1.

\bibitem{ref:ftest}
\hrefCMSnoop {}{R.~A. Fisher, ``On the interpretation of $\chi^{2}$ from
  contingency tables, and the calculation of {P}'',} \textit{ J. R. Stat. Soc.}
  \textbf{ 85} (1922) 87,
  \href{http://dx.doi.org/10.2307/2340521}{\doi{10.2307/2340521}}.

\bibitem{Dauncey:2014xga}
\hrefCMSnoop {}{P.~D. Dauncey, M.~Kenzie, N.~Wardle, and G.~J. Davies,
  ``{Handling uncertainties in background shapes}: {the discrete profiling
  method}'',} \textit{ JINST} \textbf{ 10} (2015) P04015,
  \href{http://dx.doi.org/10.1088/1748-0221/10/04/P04015}{\doi{10.1088/1748-0221/10/04/P04015}},
  \href{http://www.arXiv.org/abs/1408.6865}{\texttt{arXiv:1408.6865}}.

\bibitem{CMS-PAS-LUM-17-001}
\href {http://cdsweb.cern.ch/record/2621960}{{CMS Collaboration}, ``{CMS
  luminosity measurement for the 2017 data-taking period at $\sqrt{s}=13$
  TeV}'',} CMS Physics Analysis Summary CMS-PAS-LUM-17-001, 2018.

\bibitem{CMS-PAS-LUM-18-002}
\href {http://cdsweb.cern.ch/record/2676164}{{CMS Collaboration}, ``{CMS
  luminosity measurement for the 2018 data-taking period at $\sqrt{s}=13$
  TeV}'',} CMS Physics Analysis Summary CMS-PAS-LUM-18-002, 2019.

\bibitem{CMS:LUM-17-003}
\hrefCMSnoop {}{{CMS Collaboration}, ``Precision luminosity measurement in
  proton-proton collisions at $\sqrt{s}={13\TeV}$ in 2015 and 2016 at {CMS}'',}
  \textit{ Eur. Phys. J. C} \textbf{ 81} (2021) 800,
  \href{http://dx.doi.org/10.1140/epjc/s10052-021-09538-2}{\doi{10.1140/epjc/s10052-021-09538-2}},
  \href{http://www.arXiv.org/abs/2104.01927}{\texttt{arXiv:2104.01927}}.

\bibitem{Junk:1999kv}
\hrefCMSnoop {}{T.~Junk, ``{Confidence level computation for combining searches
  with small statistics}'',} \textit{ Nucl. Instrum. Meth. A} \textbf{ 434}
  (1999) 435,
  \href{http://dx.doi.org/10.1016/S0168-9002(99)00498-2}{\doi{10.1016/S0168-9002(99)00498-2}},
  \href{http://www.arXiv.org/abs/hep-ex/9902006}{\texttt{arXiv:hep-ex/9902006}}.

\bibitem{Read:2002hq}
\hrefCMSnoop {}{A.~L. Read, ``{Presentation of search results: The
  $\mathrm{CL}_{\mathrm{s}}$ technique}'',} \textit{ J. Phys. G} \textbf{ 28}
  (2002) 2693,
  \href{http://dx.doi.org/10.1088/0954-3899/28/10/313}{\doi{10.1088/0954-3899/28/10/313}}.

\bibitem{Cowan:2010js}
\hrefCMSnoop {}{G.~Cowan, K.~Cranmer, E.~Gross, and O.~Vitells, ``{Asymptotic
  formulae for likelihood-based tests of new physics}'',} \textit{ Eur. Phys.
  J. C} \textbf{ 71} (2011) 1554,
  \href{http://dx.doi.org/10.1140/epjc/s10052-011-1554-0}{\doi{10.1140/epjc/s10052-011-1554-0}},
  \href{http://www.arXiv.org/abs/1007.1727}{\texttt{arXiv:1007.1727}}.
  [Erratum: \DOI{10.1140/epjc/s10052-013-2501-z}].

\bibitem{DIENES1997104}
\hrefCMSnoop {}{K.~R. Dienes, C.~F. Kolda, and J.~March-Russell, ``Kinetic
  mixing and the supersymmetric gauge hierarchy'',} \textit{ Nucl. Phys. B}
  \textbf{ 492} (1997) 104,
  \href{http://dx.doi.org/10.1016/S0550-3213(97)00173-9}{\doi{10.1016/S0550-3213(97)00173-9}},
  \href{http://www.arXiv.org/abs/hep-ph/9610479}{\texttt{arXiv:hep-ph/9610479}}.

\bibitem{dyturbo}
S.~Camarda\hrefCMSnoop {}{ {et~al.}, ``{DYTurbo: Fast predictions for Drell-Yan
  processes}'',} \textit{ Eur. Phys. J. C} \textbf{ 80} (2020) 251,
  \href{http://dx.doi.org/10.1140/epjc/s10052-020-7757-5}{\doi{10.1140/epjc/s10052-020-7757-5}},
  \href{http://www.arXiv.org/abs/1910.07049}{\texttt{arXiv:1910.07049}}.

\bibitem{darkcast}
\hrefCMSnoop {}{P.~Ilten, Y.~Soreq, M.~Williams, and W.~Xue, ``{Serendipity in
  dark photon searches}'',} \textit{ JHEP} \textbf{ 06} (2018) 004,
  \href{http://dx.doi.org/10.1007/JHEP06(2018)004}{\doi{10.1007/JHEP06(2018)004}},
  \href{http://www.arXiv.org/abs/1801.04847}{\texttt{arXiv:1801.04847}}.

\end{thebibliography}\endgroup
\cleardoublepage \appendix\section{The CMS Collaboration \label{app:collab}}\begin{sloppypar}\hyphenpenalty=5000\widowpenalty=500\clubpenalty=5000
\cmsinstitute{Yerevan Physics Institute, Yerevan, Armenia}
{\tolerance=6000
A.~Hayrapetyan, A.~Tumasyan\cmsAuthorMark{1}\cmsorcid{0009-0000-0684-6742}
\par}
\cmsinstitute{Institut f\"{u}r Hochenergiephysik, Vienna, Austria}
{\tolerance=6000
W.~Adam\cmsorcid{0000-0001-9099-4341}, J.W.~Andrejkovic, T.~Bergauer\cmsorcid{0000-0002-5786-0293}, S.~Chatterjee\cmsorcid{0000-0003-2660-0349}, K.~Damanakis\cmsorcid{0000-0001-5389-2872}, M.~Dragicevic\cmsorcid{0000-0003-1967-6783}, A.~Escalante~Del~Valle\cmsorcid{0000-0002-9702-6359}, P.S.~Hussain\cmsorcid{0000-0002-4825-5278}, M.~Jeitler\cmsAuthorMark{2}\cmsorcid{0000-0002-5141-9560}, N.~Krammer\cmsorcid{0000-0002-0548-0985}, D.~Liko\cmsorcid{0000-0002-3380-473X}, I.~Mikulec\cmsorcid{0000-0003-0385-2746}, J.~Schieck\cmsAuthorMark{2}\cmsorcid{0000-0002-1058-8093}, R.~Sch\"{o}fbeck\cmsorcid{0000-0002-2332-8784}, D.~Schwarz\cmsorcid{0000-0002-3821-7331}, M.~Sonawane\cmsorcid{0000-0003-0510-7010}, S.~Templ\cmsorcid{0000-0003-3137-5692}, W.~Waltenberger\cmsorcid{0000-0002-6215-7228}, C.-E.~Wulz\cmsAuthorMark{2}\cmsorcid{0000-0001-9226-5812}
% ...
\par}
\vskip\cmsinstskip
\dag:~Deceased\\
$^{1}$Also at Yerevan State University, Yerevan, Armenia\\
$^{2}$Also at TU Wien, Vienna, Austria\\
% ...
\end{sloppypar}
\end{document}